\documentclass[12pt,preprint]{aastex}
\shorttitle{Galaxy Spins in cDE Models}
\begin{document}
\title{The Spin Alignments in Galaxy Pairs as a Test of Bouncing Coupled Dark 
Energy}
\author{Jounghun Lee}
\affil{Department of Physics and Astronomy, Seoul National University, Seoul 
151-747, Korea}
\email{jounghun@astro.snu.ac.kr}
\begin{abstract}
We investigate the effect of coupled dark energy (cDE) on the spin alignments 
in isolated pairs of galactic halos, using the publicly available data from the 
hydrodynamical cDE simulations (H-CoDECs) which were run for various cDE models 
such as  EXP001, EXP002, EXP003 (with exponential potential and constant 
coupling), EXP008e3 (with exponential potential and exponential coupling) and 
SUGRA003 (with supergravity potential and negative constant coupling) 
as well as for a standard $\Lambda$CDM cosmology (with the WMAP7 parameters). 
Measuring the cosines of the angles between the spin axes in isolated pairs of 
galactic halos for each model and determining its probability density 
distribution, we show that for the SUGRA003 model with bouncing cDE the null 
hypothesis of no spin alignment in pairs of galactic halos is rejected at 
$99.999\%$ confidence level. In contrast, the $\Lambda$CDM cosmology 
yields no significant signal of spin alignment and the other four cDE 
models also exhibit only weak signals of spin-alignments.
The strength of the spin alignment signal is found to be almost independent 
of the total halo mass and separation distance in galaxy pairs.
Showing also that no signal is detected from the Sloan Digital Sky Survey 
DR 7,  we conclude that the spin alignments in galaxy pairs is in principle a 
unique test of bouncing cDE models.
\end{abstract}
\section{INTRODUCTION}

Recent observations have warned us that the $\Lambda$CDM ($\Lambda$+cold 
dark matter) cosmology might not be the ultimate truth of the Universe even 
though it works well on the large scale. 
The increasing amount of the observational evidences on the (sub-)galactic scale 
against the $\Lambda$CDM model that have been accumulated for the past decade 
should be no longer dismissed nor overlooked under the shield of unknown 
complicated baryonic physics \citep[see e.g.,][and references therein]
{puzzle,deblok10,PN10,kroupa-etal10,kuzio-etal11}. 

The mismatches of the theoretical predictions based on the $\Lambda$CDM model 
with the observations on the (sub-)galactic scales have so far been 
routinely attributed to our lack of knowledge about all complicated baryonic 
processes involved in the formation of galaxies. 
But the recent rapid progresses in observational and numerical studies  
have indicated that it may not fully rescue the $\Lambda$CDM cosmology to take 
into account highly nonlinear baryon effects. For example, the observed flat 
density cores of the dark-matter-dominated low surface brightness galaxies 
(LSBGs) \citep[see][for a comprehensive review]{deblok10} have been regarded 
as one of the most serious observational challenges on the sub-galactic scale to 
the $\Lambda$CDM cosmology which predicts the cuspy density cores \citep{nfw96}. 
Recent work of \citet{kuzio-etal11} have revealed that the baryonic processes 
are unlikely to flatten the density cores of the LSBGs, by comparing thoroughly 
the simulated disk galaxies in hydrodynamical simulations with the observed 
LSBGs.

Nevertheless, there is a good reason that the $\Lambda$CDM model is still 
regarded as the standard one. Although many authors have so far made strenuous 
endeavors to devise a better (and hopefully more fundamental) cosmological theory 
expecting it to work as well as the $\Lambda$CDM model on the large scale 
while simultaneously resolving all of the (sub-)galactic scale tensions of the 
$\Lambda$CDM cosmology with observations, no alternative theory has so far 
been capable of defeating the $\Lambda$CDM model. 
The difficulty in coming up with a viable alternative lies in the fact that the 
small-scale agreements of a model tends to be achieved only at the cost of its 
large-scale agreements and vice-versa. To make matters worse, when a model 
was found to be consistent with the observations on the large scales, then 
it has turned out to be not distinguishable from the $\Lambda$CDM cosmology. 

Unprecedentedly ample amount of high-quality data recently available from 
observations and hydrodynamical simulations, however, gives us a hope that 
it might be possible to eventually distinguish between the candidate cosmological 
models and to finally rule some of them out. The key question is what statistical 
property of the Universe on the (sub-)galactic scale would be the most useful 
tool to achieve this goal.
Here, we suggest the spin alignments in isolated galaxy pairs as a unique tool to 
test coupled dark energy (cDE) models, one of the currently popular alternatives 
to the $\Lambda$CDM cosmology, in which the acceleration of the Universe is 
driven by a dynamical scalar field dark energy (DE), $\phi$, coupled to the 
non-baryonic CDM under the influence of scalar self-interaction potential 
$U(\phi)$ \citep{RP88,wetterich95,amendola00,amendola04}. 

In the cDE picture the long-range fifth force generated by the DE-CDM coupling 
plays a role of enhancing the density and velocity perturbations in the linear 
regime relative to the $\Lambda$CDM case \citep{baldi11b,codecs,LB11}, 
which eventually leads to the earlier formation and faster merging of 
dark halos \citep[][and references therein]
{mangano-etal03,maccio-etal04,MB06,PB08,baldi-etal10,WP10,BV10,sugra003}.
Since the galaxy angular momentum depends strongly on both of the density 
and velocity perturbations in the linear regime, 
the relative spin orientations in galaxy pairs would differ among the candidate 
cDE models which are characterized by different shapes of the coupling function 
$\beta(\phi)$ and self-interaction potential $U(\phi)$.

The following two recent literatures have motivated us to do this work. The first 
one is \citet{cervantes-etal10} who showed that there is no significant spin 
alignments in pairs of late-type galaxies from the seventh data release of 
the Sloan Digital Sky Survey \citep[][hereafter, SDSS DR7]{sdssdr7}. 
The second one is \citet{codecs} who performed high-resolution hydrodynamic 
simulations (H-CoDECs) for various cDE models and released the data to the public 
very recently. Analyzing the H-CoDECs halo catalogs at $z=0$, we find it possible 
in principle to test the cDE models by comparing their predictions on the spin 
alignments in galaxy pairs with the observational result.

The outline of this Paper is as follows. In \S 2, a brief description of the 
halo and subhalo catalogs from H-CoDECS is provided. In \S 3, we explain how 
the spin alignments in isolated pairs of galactic halos identified using H-CoDECS 
data are determined, how the probability density distribution of the alignment 
angles is derived at $z=0$ for each cDE model, and how the spin alignment signals 
depend on the total mass and separation distance in galaxy pairs. 
In \S 4, the spin alignments in pairs of late-type galaxies from the SDSS DR7 
are calculated and its probability density distribution is compared with the 
numerical results from the H-CoDECS. In \S 5, the possibility of testing bouncing 
cDE model with the spin alignments in galaxy pairs is discussed and a final 
conclusion is stated.

\section{DATA FROM THE H-CODECS}

\citet{codecs} conducted a series of H-CoDECS (Hydrodynamical Coupled Dark Energy 
Simulations) with the modified GADGET-2 code \citep{gadget2,baldi-etal10} on a 
periodic box of volume $0.512\,h^{-3}{\rm Gpc}^{3}$, which contain approximately 
$134217728$ gas particles on the top of the same number of CDM particles. 
The individual gas and CDM particle mass is 
$M_{\rm b}=4.78\times 10^{7}\,h^{-1}M_{\odot}$ and  
$M_{\rm CDM}=2.39\times 10^{8}\,h^{-1}M_{\odot}$ at $z=0$, respectively. 
Using the SPH (Smoooth Particle Hydrodynamics) technique incorporated in the 
GADGET-2 code \citep{baldi-etal10}, the adiabatic hydrodynamical forces 
were computed in each H-CoDECS runs to track down the dynamical evolution of 
the gas particles.

Released to the public were all the numerical data from the H-CoDECS for five 
different cDE models, namely,  EXP001, EXP002, EXP003, EXP008e3 and SUGRA003 
as well as the standard $\Lambda$CDM model. The simulation runs for all 
six models started from the same initial conditions consistent with the WMAP7 
parameters and the linear power spectra of all models are also normalized to 
have the same amplitudes at the epoch of decoupling. The first three cDE models 
have the exponential form of the scalar self-interaction potential 
\citep{LM85,RP88,wetterich88}, $U(\phi)\propto e^{-0.08\phi}$, and constant 
coupling functions: $\beta=0.05,\ 0.1$ and $0.15$ for the EXP001, EXP002 and 
EXP003, respectively, which are all within the current limits on the cDE 
coupling strength from observations \citep{bean-etal08,xia09,BV10,BS11}. 

The EXP008e3 model has the same exponential potential, 
$U(\phi)\propto e^{-0.08\phi}$, but with time-dependent coupling function of  
$\beta=0.4e^{3\phi}$ \citep{exp008e3}. 
On the other hand, the SUGRA003 is a bouncing cDE model characterized by 
supergravity potential \citep{BM99}, 
$U(\phi)\propto\phi^{-\alpha}e^{\phi^{2}/2}$, and by negative value of 
constant coupling of $\beta=-0.15$. See \citet{codecs} who provides much more 
comprehensive introduction of general cDE cosmologies as well as more detailed 
explanations on the above six specific cDE models 
\citep[see also][and references therein]
{amendola00,amendola04,PB08,baldi-etal10,BLM11,exp008e3,sugra003}.

The bound halos and their subhalos are identified by applying the standard 
Friends-of-Friends (FoF) with linkage parameter of $0.2$) and the SUBFIND algorithm 
to the H-CoDECS, respectively, for each cosmological model \citep{FoF02,gadget2}.
The H-CoDECS halo catalog provides information on the number of subhalos ($N_{s}$), 
FoF mass, position, and velocity of each halo, while information on the numbers of particles 
($N_{p}$), masses ($m$), positions, specific angular momenta ($\hat{\bf J}$) of 
all the subhalos belonging to each halo is available in the H-CoDECS substructure catalog. 
Here, the halo mass corresponds to the collapsed mass measured at 
$z=0$ as the sum of all the gas and CDM particles belonging to the same 
FoF groups. Thus the halo mass has the same meaning in the six cosmological models 
since the mass of each individual particle has the same value at $z=0$ in all of 
the six models \citep[e.g., see][]{sugra003}. 
For more detailed explanations on the H-CoDECS project and its halo/substructure 
catalogs, see \citet{codecs} and visit the CoDECS webpage
\footnote{It is http://www.marcobaldi.it/web/CoDECS.html}. 

\section{SPIN ALIGNMENTS IN ISOLATED PAIRS OF GALACTIC HALOS}

Analyzing the halo and subhalo catalogs at $z=0$ from H-CoDECS for each 
cosmological model, we identify those halos which have only two subhalos 
($N_{s}=2$) without belonging  to any larger halos and regard the two subhalos 
in each identified halo as an isolated galaxy pair. To avoid those poorly 
resolved subhalos whose specific angular momentum vectors are likely to suffer 
from inaccurate measurements, we select only those halos among the identified 
ones in which each of the two subhalos has $100$ or more particles. 
We focus here only on the {\it isolated} pairs of galactic halos located in 
low-density regions to control the environmental effect on the spin 
orientations to the minimum level, which help single out the cDE effect. 

Table \ref{tab:pair} lists the number ($N_{pair}$) of the selected galaxy pairs, the 
median mass of the halo ($M_{\rm T, med}$), the median masses of the two 
subhalos ($m_{\rm 1, med}$ and $m_{\rm 2,med}$ in a decreasing order) in 
pairs for the six cosmological models.  As can be seen, the six samples have almost 
the same median masses. The maximum difference in $M_{\rm T, med}$ and in 
$m_{\rm i, med}$ among the six models is less than $10\%$. 

\begin{figure}[ht]
\begin{center}
\plotone{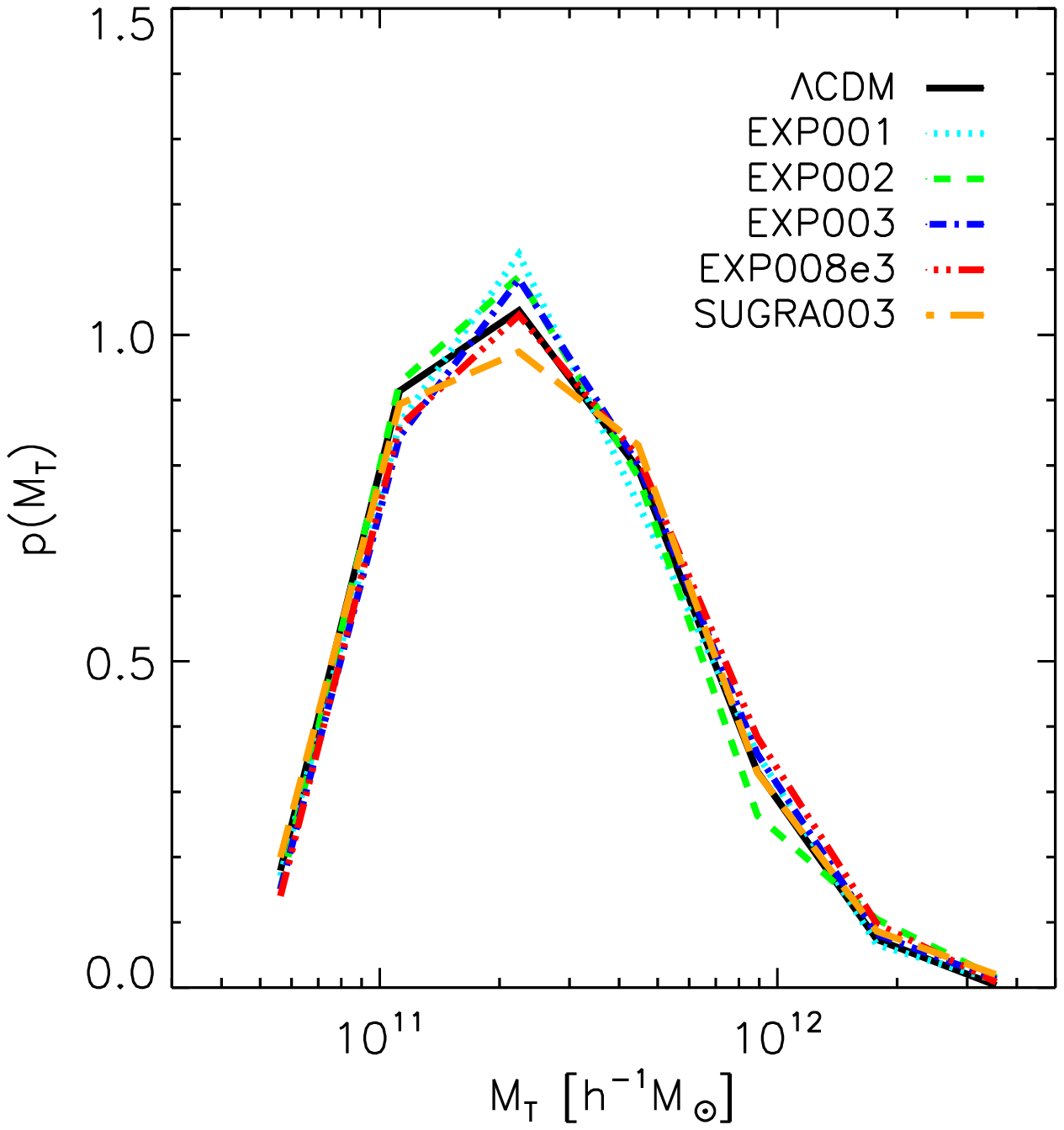}
\caption{Probability density distributions of the total mass of the galaxies in pairs 
for the six models.}
\label{fig:m_dis}
\end{center}
\end{figure}

\begin{figure}
\begin{center}
\plotone{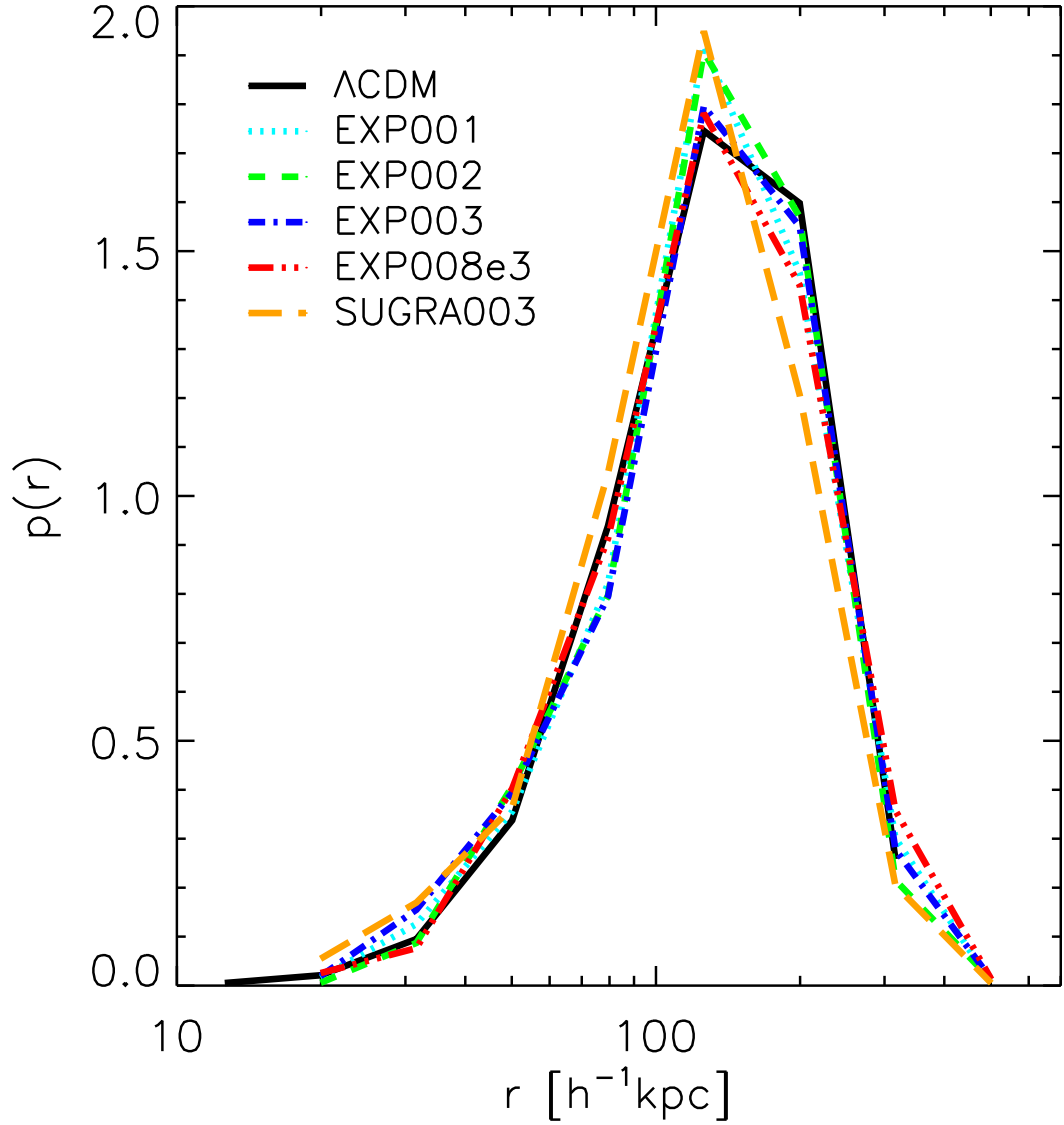}
\caption{Probability density distributions of the separation distance between the 
galaxies in pairs for the six models.}
\label{fig:r_dis}
\end{center}
\end{figure}

Figure \ref{fig:m_dis} plots the probability density distributions of the total mass of the 
selected galaxy pairs, $p(M_{\rm T})$, for the six cosmological models. 
As can be seen, the distributions $p(M_{\rm T})$ are very similar to one another with 
almost the same width, height, and location of the maximum value of $p(M_{\rm T})$ 
at ($3\times 10^{11}\,h^{-1}M_{\odot}$) among the six models. 
This result confirms that our six samples of the isolated galaxy pairs are comparable to 
one another in mass. In fact,  this result should be naturally expected since the same algorithm 
was employed to find the isolated galaxy pairs for each model case. 
Figure \ref{fig:r_dis} plots the probability density distributions of the separation distance 
in galaxy pair, $p(r)$. As can be seen, the distributions are quite similar to one another, 
too, having the characteristic small-$r$ tail, reaching their maximum values 
at $r\approx 150\,h^{-1}$kpc. 

For each selected galaxy pair, we calculate the cosine of the angle, 
$\cos\theta$, between the two specific angular momentum vectors as 
$\cos\theta\equiv\vert\hat{\bf J}_{1}\cdot\hat{\bf J}_{2}\vert
/(\hat{J}_{1}\hat{J}_{2})$ where $\hat{J}_{i}\equiv\vert\hat{\bf J}_{i}\vert$.
Here we restrict the range of the angle, $\theta$, 
to $[0,\ \pi/2]$ since what matters is not the signs of the 
two angular momentum vectors but their relative orientations. 
Binning the values of $\cos\theta$ in range of $[0,\ 1]$ and counting the number 
of the selected galaxy-pairs belonging to each $\cos\theta$-bin, we determine the 
probability density distribution, $p(\cos\theta)$, for each cosmological 
model. If the two specific angular momentum vectors in galaxy pairs have strong 
tendency to be aligned with each other, then the probability distribution 
$p(\cos\theta)$ is expected to increase as $\cos\theta$ increases. If there is 
no alignment tendency, then it must be uniform, $p(\cos\theta)=1$, 
over the range of $0\le \cos\theta\le 1$. The stronger alignment tendency will 
be manifest as the higher degree of the deviation of $p(\cos\theta)$ from the 
uniform distribution. 

\begin{figure}
\begin{center}
\plotone{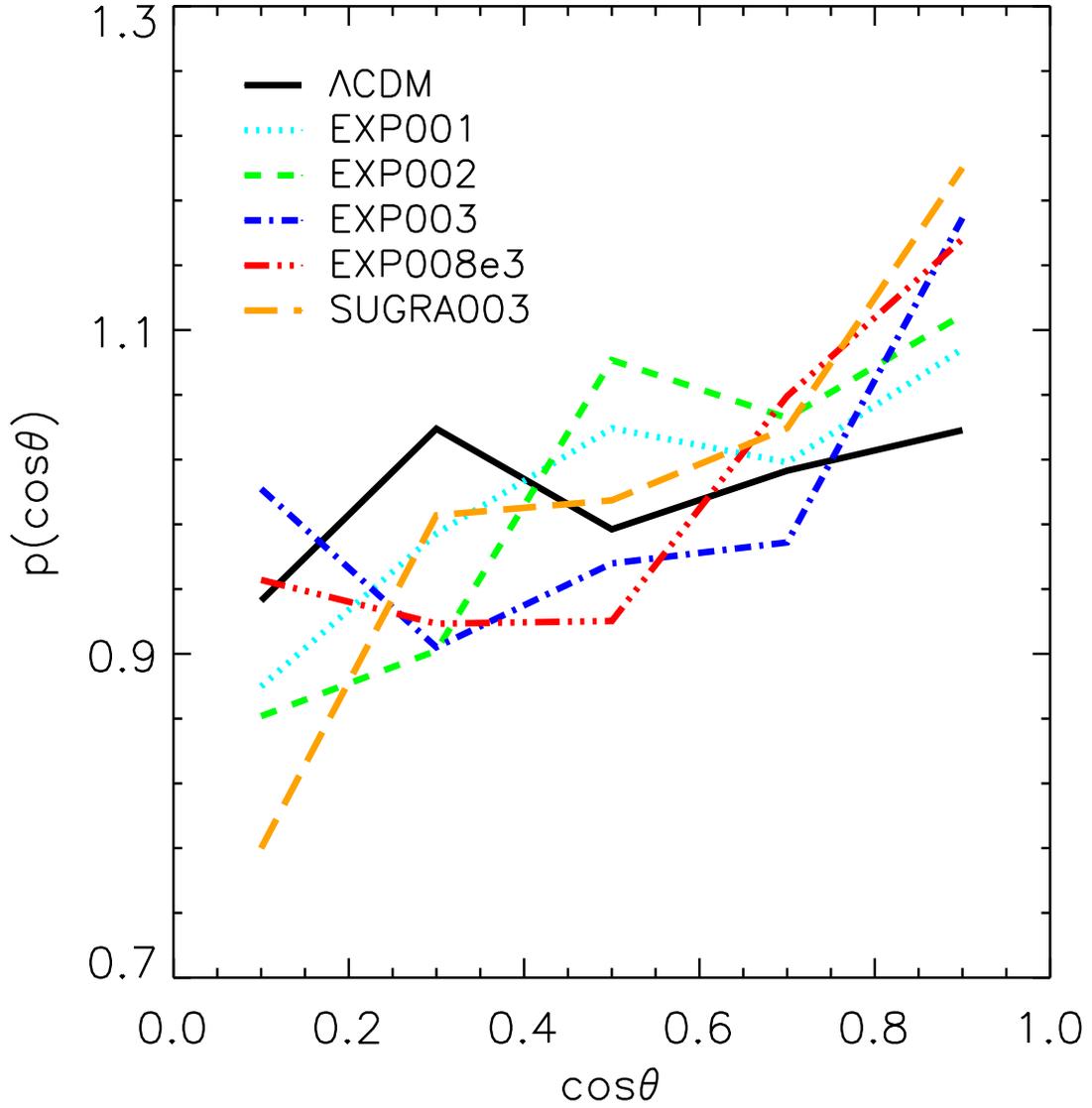}
\caption{Probability density of the cosines of the angles between the spin 
axes of the galaxies in pairs for the six models. }
\label{fig:cost_dis}
\end{center}
\end{figure}

Figure \ref{fig:cost_dis} plots the probability density distributions of $\cos\theta$ 
for the six models. As can be seen, for the $\Lambda$CDM case, the probability density 
$p(\cos\theta)$ is almost uniform. For the case of the constant coupling, the larger the 
coupling constant $\beta$ is, the more severely the probability density 
$p(\cos\theta)$ seems to deviate from the uniform distribution. 
Nevertheless, even for the extreme case of EXP003 (with $\beta=0.15$) the degree of 
the deviation of $p(\cos\theta)$ from the uniform distribution is not so high as in the 
SUGRA003 model which exhibits the highest degree of the deviation. 

\begin{figure}
\begin{center}
\plotone{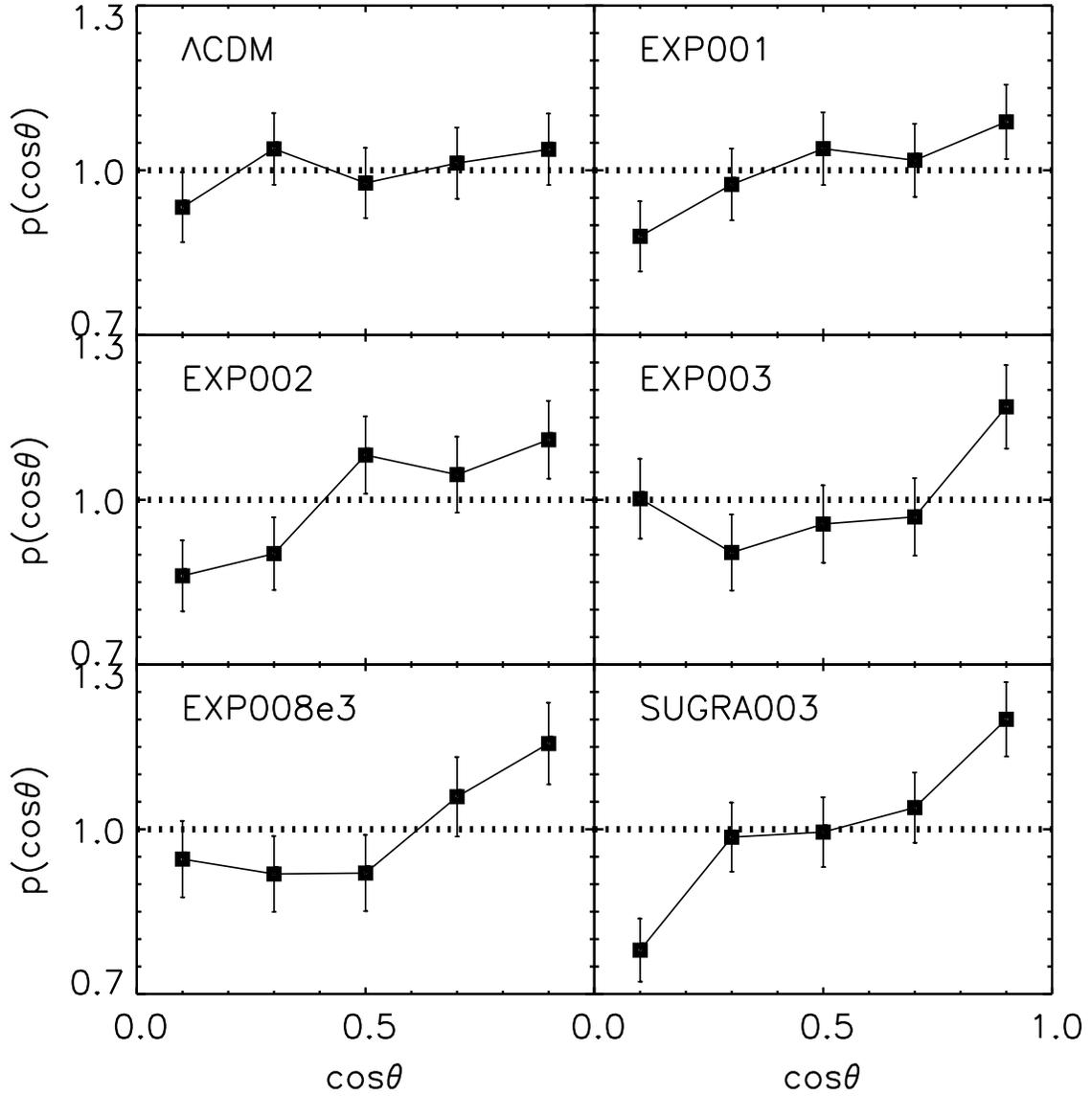}
\caption{Same as Figure \ref{fig:cost_dis} but with the bootstrap errors that are 
calculated as one $\sigma$ scatter among the $10000$ bootstrap resamples in 
the six separate panels.}
\label{fig:boot_err}
\end{center}
\end{figure}

To estimate the statistical significance of the alignment signals detected for the 
cDE model cases, we perform the bootstrap error analysis.  Figure \ref{fig:boot_err} 
plots the same as Figure \ref{fig:cost_dis} but in the separate panels, showing the 
bootstrap errors, 
$\sigma_{\rm boot}$, calculated as one standard deviation of $\cos\theta$ among the 
$10000$ bootstrap resamples. In each panel the horizontal dotted line corresponds to the 
uniform probability density for the case of no alignment. As can be seen, for the 
$\Lambda$CDM case, $p(\cos\theta)$ is almost perfectly consistent with the 
uniform distribution.  For the SUGRA003 case, the alignment signal is as significant as 
$4\sigma_{\rm boot}$ in the first and the fifth bin. For the other cDE model cases, the 
alignment signal is not so significant as the SUGRA003 case.

\begin{figure}[ht]
\begin{center}
\plotone{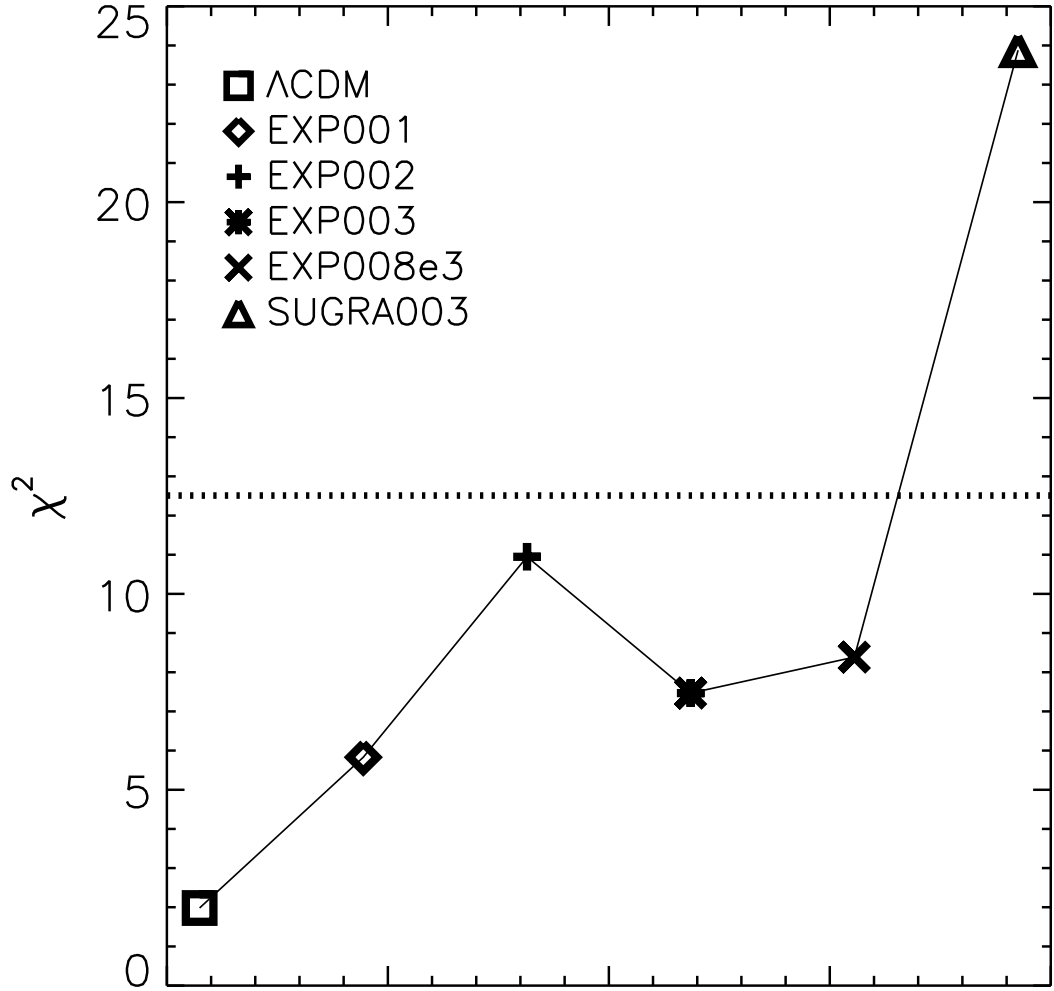}
\caption{$\chi^{2}$ values (see Eq.[\ref{eqn:chi2}]) for the six models.
The dotted horizontal line indicates the value of $\chi^{2}$ which 
leads to the rejection of the null hypothesis at the $95\%$ confidence level. }
\label{fig:chi2}
\end{center}
\end{figure}

To test the null hypothesis of $p(\cos\theta)=1$, we also calculate $\chi^{2}$ with 
$N_{\rm bin}-1$ degree of freedom as:
\begin{equation}
\chi^{2}=\sum_{i=1}^{N_{\rm bin}}
\frac{[p(\cos\theta_{i})-1]^{2}}{\sigma^{2}_{\rm boot}},
\label{eqn:chi2}
\end{equation}
where $N_{\rm bin}$ is the number of the $\cos\theta$-bin and $p(\cos\theta_{i})$ 
represents the value of the probability density distribution at the $i$-th $\cos\theta$ bin. 
Figure \ref{fig:chi2} plots the values of $\chi^{2}$ versus the models. The horizontal 
dotted line corresponds to the value of $\chi^{2}$ with which the null hypothesis is 
rejected at the $99\%$ confidence level \citep{WJ03}. 
It is found that for the SUGRA003 model the null hypothesis of no spin 
alignment in the galaxy pairs is rejected at the $99.999\%$ confidence level. 

\begin{figure}[ht]
\begin{center}
\plotone{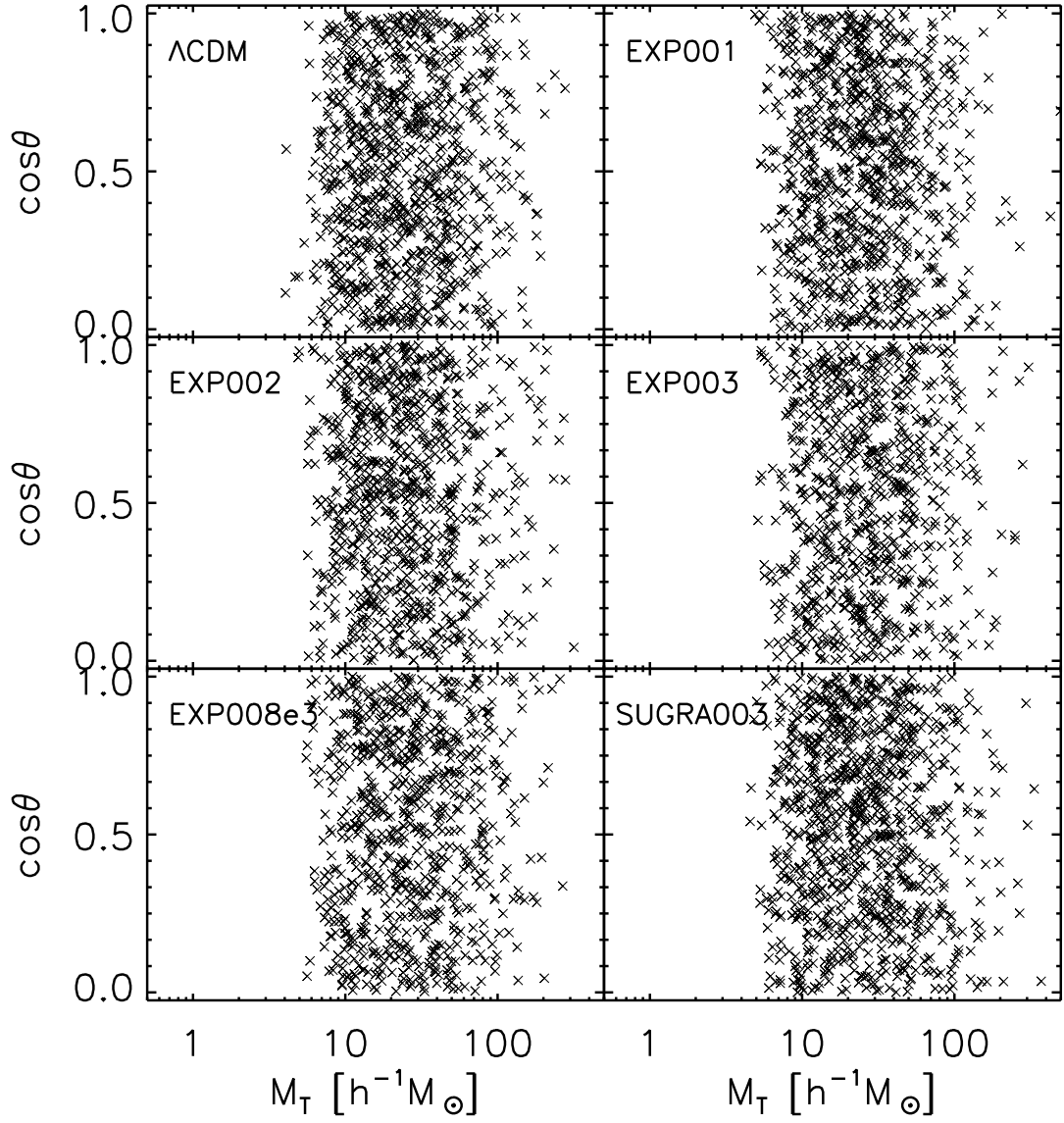}
\caption{Scatter plots of the cosines of the alignment angles versus the total 
masses in isolated galaxy pairs for the six models.}
\label{fig:sca_m}
\end{center}
\end{figure}

\begin{figure}[ht]
\begin{center}
\plotone{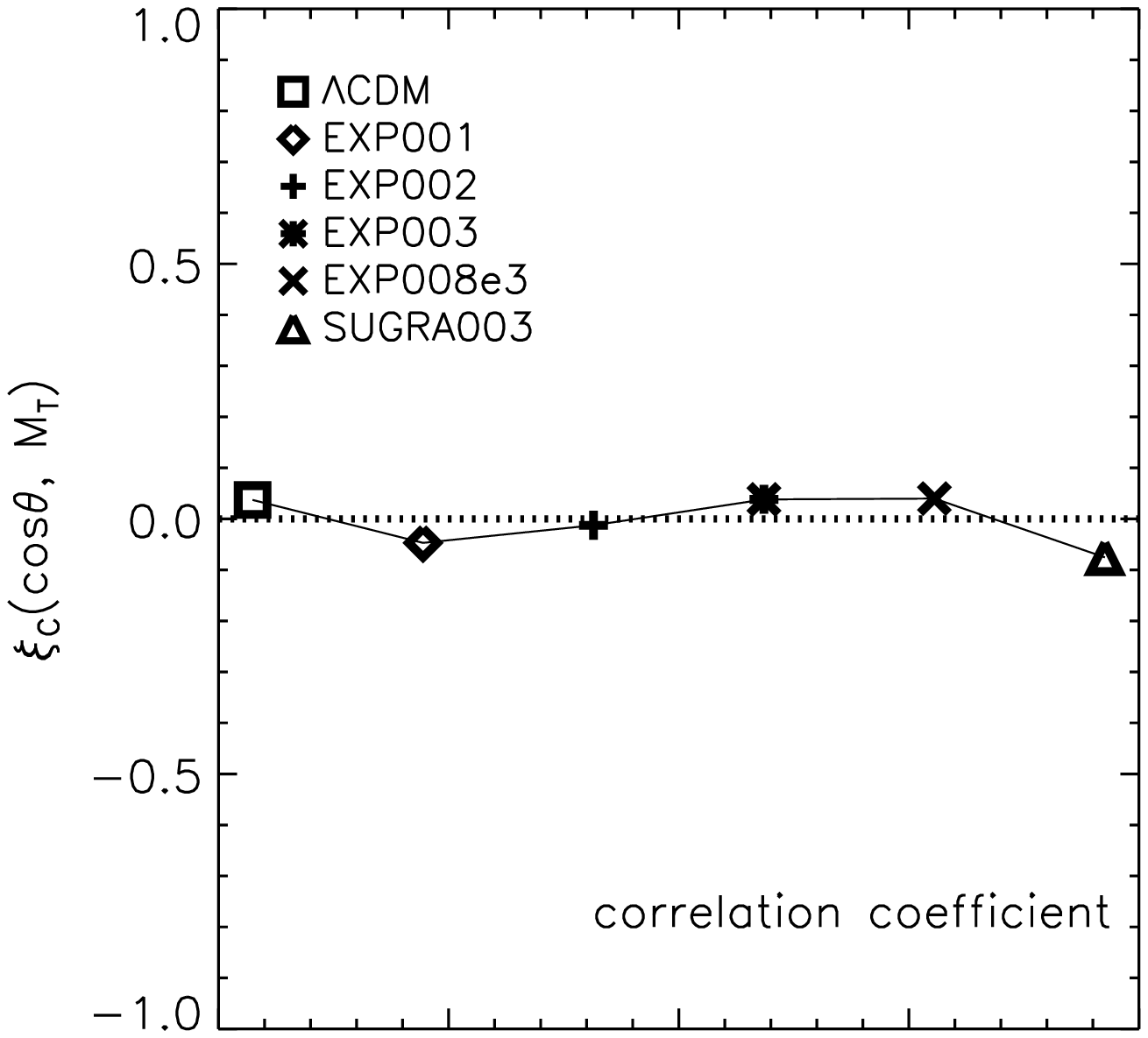}
\caption{Correlation coefficient of $\cos\theta$ and $M_{\rm T}$ for the 
six models. }
\label{fig:xi}
\end{center}
\end{figure}

Now, we would like to investigate if there is any dependence of the spin alignment 
signal on the mass $M_{T}$. This issue may be important to address since it has been 
recently reported that the degree of the shape alignments of the galaxy groups with 
the large-scale structures depends on the mass scale \citep{paz-etal11}.
Figure \ref{fig:sca_m} plots $\cos\theta$ versus 
$M_{\rm T}$ for the six models in the separate panels. As can be seen, there exists 
no obvious correlation between $\cos\theta$ and $M_{\rm T}$. To address more 
this issue quantitatively, we calculate the correlation coefficient, $\xi$, of 
$\cos\theta$ and $M_{\rm T}$ as \citep{WJ03}
\begin{equation}
\xi(\cos\theta,M_{\rm T} )= \frac{\langle (\cos\theta-\langle\cos\theta\rangle)
(M_{\rm T}-\langle M_{\rm T}\rangle)\rangle}
{\left[\langle (\cos\theta-\langle\cos\theta\rangle)^{2}\rangle
\langle(M_{\rm T}-\langle M_{\rm T}\rangle)^{2}\rangle\right]^{1/2}},
\label{eqn:xi}
\end{equation}
where the average is taken over all the selected galaxy pairs and the value of $\xi$ 
ranges between $-1$ and $1$. 
If there is a strong correlation (anti-correlation) between $\cos\theta$ and $M_{\rm T}$, 
then $\xi$ will be close to $1$ ($-1$). Whereas, if there is no correlation between the two 
quantities, $\xi$ will be zero. The weaker the correlation between $\cos\theta$ and 
$M_{\rm T}$ is, the closer to zero the value of $\xi$ becomes. Figure \ref{fig:xi} plots 
$\xi$ versus the models. As can be seen, for each model the value of $\xi$ is less then 
$0.1$, which clearly demonstrates  that the spin alignment signal in a isolated galaxy pair 
hardly depends on its total mass.

\begin{figure}[ht]
\begin{center}
\plotone{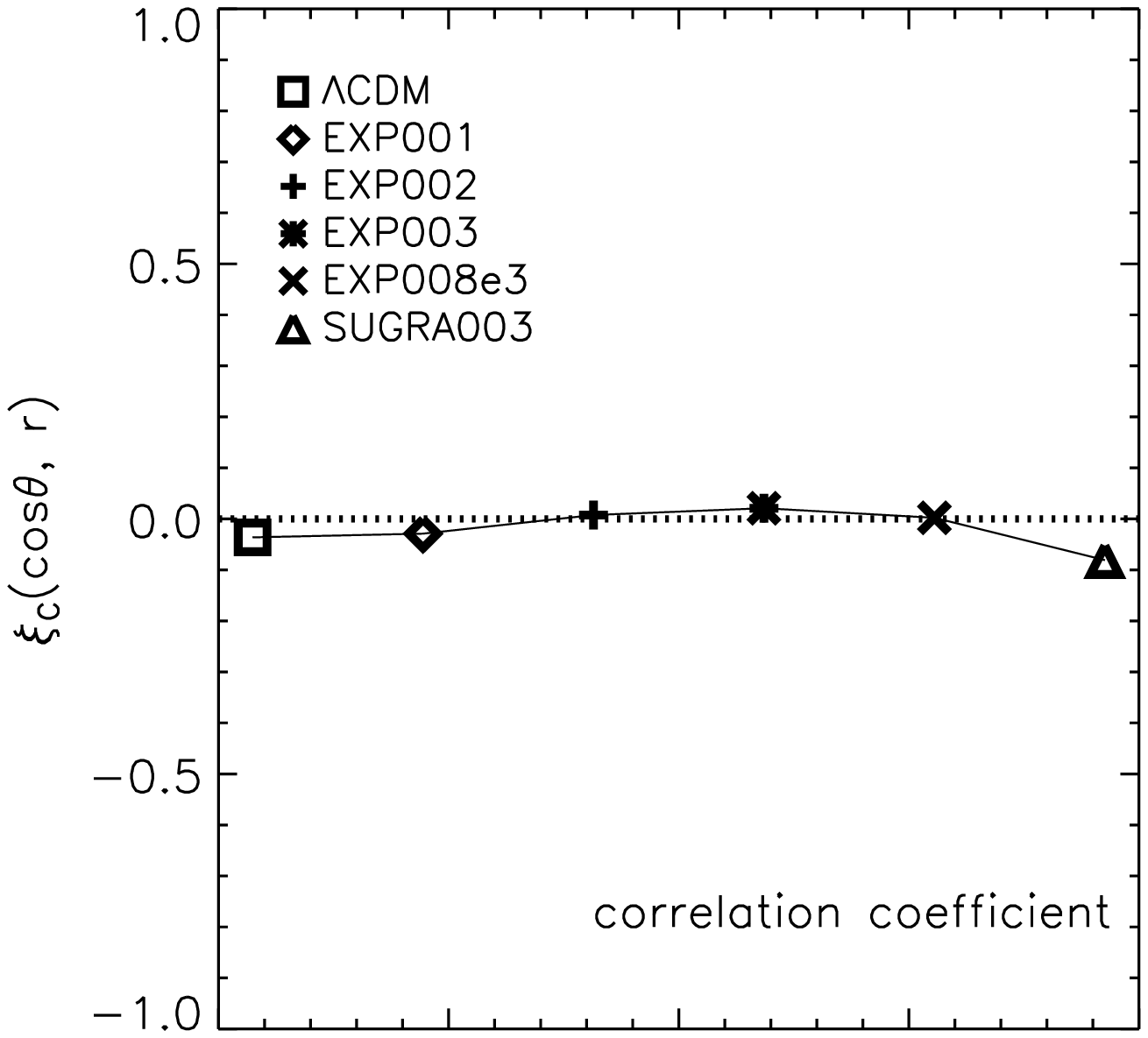}
\caption{Correlation coefficient of $\cos\theta$ and $r$ for the 
six models.}
\label{fig:xi_r}
\end{center}
\end{figure}

Similarly, we examine if there is any correlation between $\cos\theta$ and the 
separation distance between the pair galaxies, $r$. 
The values of the correlation coefficient, $\xi(\cos\theta,r)$, are calculated through 
substituting $r$ for $M_{T}$ in Equation (\ref{eqn:xi}) for the six models, the results 
of which are plotted in Figure \ref{fig:xi_r}. As can be seen, the value of $\xi$ is 
very close to zero, confirming that there is basically no correlation between 
$\cos\theta$ and $r$.

\section{COMPARISON WITH OBSERVATION FROM SDSS DR7}

As mentioned in \S 2, it was \citet{cervantes-etal10} who have for the first 
time shown that no significant degree of alignment exists between the spin axes 
in galaxy pairs, analyzing the $255$ pairs of late-type galaxies in redshift range of 
$[0.01,\ 0.2]$ identified using a spectroscopic sample from the SDSS DR7. 
Intriguing as their result may be, their analysis definitely has some room for 
improvement. For example, they assumed naively that the angle between the 
spin axes in a galaxy pair equals the difference between the position 
angles, which is not necessarily true. Besides, considering a relative 
wide redshift interval to find galaxy pairs, they measured the relative 
spin orientations without accounting for the fact that the spin axes of those spiral 
galaxies located at $z\ge 0.05$ are difficult to measure accurately due to the 
systematics caused by the presence of bulges \citep{lee11}.

In this section, we present a more robust and thorough analysis for the measurements 
of the spin alignments in pairs of the SDSS late-type galaxies. First of all,
we utilize the spectroscopic catalog of the SDSS DR7 galaxies compiled by 
\citet{huertas-etal11}, in which a total of $698420$ galaxies at $0\le z\le 0.16$ 
are all classified by means of the Bayesian statistics into five Hubble types, 
\{ E, Ell, S0, Sab Scd\} and each of them is assigned corresponding 
five probabilities P(E), P(Ell), P(S0), P(Sab), P(Scd). 
Out of this catalog, we construct a sample for our analysis, selecting only 
those nearby large late-type galaxies which satisfy the conditions of 
$z\le 0.02$, $D\ge 7.92$ $arc sec$ and 
P(Scd)=max\{P(E),P(Ell),P(S0),P(Sab),P(Scd)\} where $D$ denotes 
the diameter of a given late-type galaxy \citep{lee11}. That is, we focus only on 
the nearby large late-type galaxies since their spin axes are relatively accurately 
measurable owing to their small bulges and large extended disks \citep{lee11}.

Taking each galaxy from the sample as a target, we find its nearest neighbor 
galaxy in the same sample. Let $d_{s}$ and $\Delta z$ be the separation distance 
and redshift difference between a target and its nearest neighbor, respectively. 
If the nearest neighbor galaxy has no galaxy within the distance of $d_{s}$ other 
than the target galaxy and if $d_{s}\le 1\,h^{-1}$Mpc and $\Delta z\le 0.001$, then 
the target galaxy and its nearest neighbor are regarded as an 
isolated pair system composed of the SDSS late-type galaxies. 

A total of $84$ pairs of the late-type galaxies are identified from our sample 
and the directions of the two spin axes in each galaxy pair is determined 
up to two-fold ambiguity (accounting for both the clock-wise and counter-clock 
wise spinning) with the help of the circular thin disk approximation as 
\citep{PLS00,LE07,lee11}
\begin{eqnarray}
\hat{J}_{x}&=&\pm\cos I\cos\delta\cos\alpha + 
\sqrt{1-\cos^{2}I}\sin I\sin\delta\cos\alpha  - 
\sqrt{1-\cos^{2}I}\cos P\sin\alpha,\\
\hat{J}_{y} &=&\pm\cos I\cos\delta\sin\alpha  + 
\sqrt{1-\cos^{2}I}\sin P\sin\delta\sin\alpha + 
\sqrt{1-\cos^{2}I}\cos P\cos\alpha,\\
\hat{J}_{z} &=& \pm\cos I\sin\delta - 
\sqrt{1-\cos^{2}I}\sin P\cos\delta, 
\end{eqnarray}
where $I$ is the inclination angle, $P$ is the position angle and 
$(\alpha,\ \delta)$ are the right ascension and declination of each 
late-type disk galaxy in pair. The inclination angle of each late-type galaxy 
is determined as $\cos^{2}I=(q^{2}-u^{2})/(1-u^{2})$ where $q$ is the 
axial ratio and $u$ is the intrinsic flatness parameter introduced by 
\citet{HG84}. Following the Bayesian approach, the value of $u$ for the 
selected late-type galaxies are calculated as  
$u=u_{Sa}{\rm P(Sa)}+u_{Sbc}{\rm P(Sbc)}+u_{Scd}{\rm P(Scd)}$ with 
$u_{Sa}=0.23,\ u_{Sbc}=0.2,\ u_{Scd}=0.1$. For a detailed 
description of the measurements of the minor axes of the spiral 
galaxies with intrinsic parameter, see \citet{HG84} and \citet{lee11}. 

Using the same procedure described in \S 3, we derive the probability density 
distribution of the cosines of the angles between the two spin axes in the 
selected galaxy pairs. Note that the two-fold ambiguity in the measurement  
of $\hat{\bf J}$ produces the four-fold ambiguity in the determination of 
$\cos\theta$. In other words, for each galaxy pair, we end up having 
four different values of $\cos\theta$. Regarding the four values as 
different realizations of the alignment angles for each galaxy pair, 
we calculate the probability density distributions of $p(\cos\theta)$ 
using the $336(=84\times 4)$ realizations of the spin alignment. 

\begin{figure}[ht]
\begin{center}
\plotone{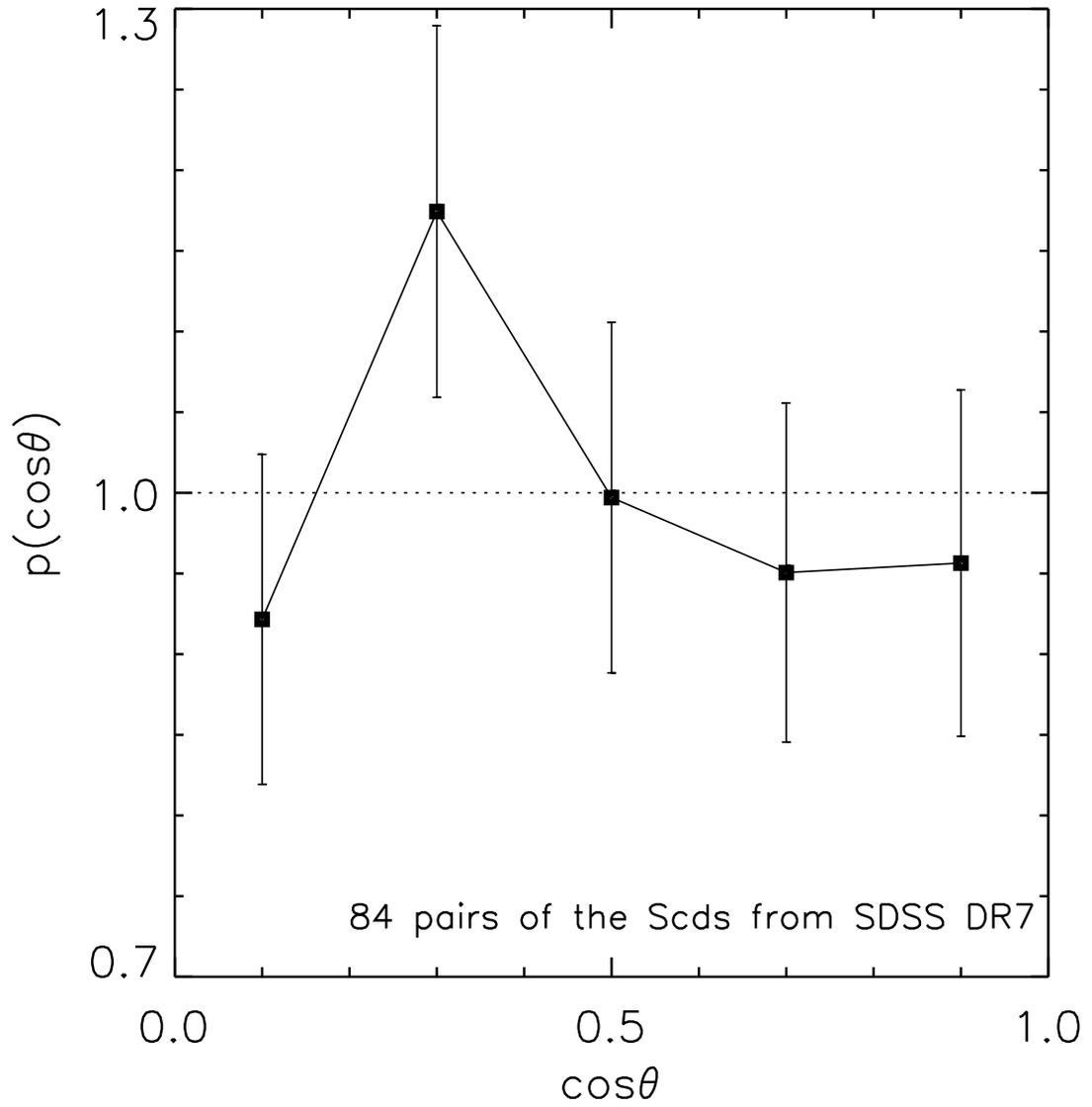}
\caption{Probability density distribution of the cosines of the angles between the 
spin axes in the 84 Scd galaxy pairs selected from the SDSS DR7. The errors are 
calculated as one $\sigma$ scatter among the 1000 bootstrap resamples.}
\label{fig:sdss}
\end{center}
\end{figure}

The result is plotted as square dots in Figure \ref{fig:sdss}. 
The errors are again calculated as one standard deviation scatter among 
$10000$ bootstrap resamples. As can be seen, the observed probability 
density distribution, $p(\cos\theta)$, is almost perfectly uniform, indicating 
no spin alignments in pairs of the late-type galaxies from SDSS DR7, which is 
consistent with the result of \citet{cervantes-etal10}. 

When this observational result is taken at its face value, it is inconsistent with the SUGRA003 
model case which yields significant signal of the spin alignments in isolated galaxy pairs.  
An acute reader, however,  would think that the observational sample of the SDSS galaxy pairs 
is biased toward higher masses than the numerical sample from H-CoDECS since the former 
includes only those large spiral galaxy pairs, while the numerical samples include all 
isolated pairs of the galaxies.  Yet, we have already shown in \S 3 that there is 
very little, if any, correlation between the spin alignment signal and the total halo 
mass in galaxy pair system. We think that although our observational sample is biased, 
this systematics is unlikely to contaminate severely the spin alignment signal. In other 
words, the result of no spin alignment from the SDSS galaxy pair sample is unlikely 
to be due to the systematics caused by including selectively only those large spiral 
galaxies (which must be biased toward large halo mass) in the sample. 

There is another difference in the observational data analysis from dealing with 
the numerical data.  The measurement of the alignment angle between the spin 
directions in each SDSS galaxy pair unavoidably suffers from the four-fold ambiguity, 
which effectively adds to $p(\cos\theta)$, lowering the significance of spin alignment 
signal.  Here, we would like to examine whether or not the spin alignment signal seen 
in the SUGRA003 model remains significant when the same assumption is applied to 
the model. Reversing the signs of the radial components of the unit spin vectors of 
the two galaxies, we construct two new unit spin vectors per each galaxy pair. 
Using the two unit spin vectors (the original one and the newly constructed one 
through reversing the sign of the radial component of the original spin vector) of 
each galaxy in pair, we calculate four times the cosines of the spin alignment angles 
per each galaxy pair. Then, we  redetermine $p(\cos\theta)$, regarding the four 
values of $\cos\theta$ per each galaxy pair as four different realizations.  
 
\begin{figure}[ht]
\begin{center}
\plotone{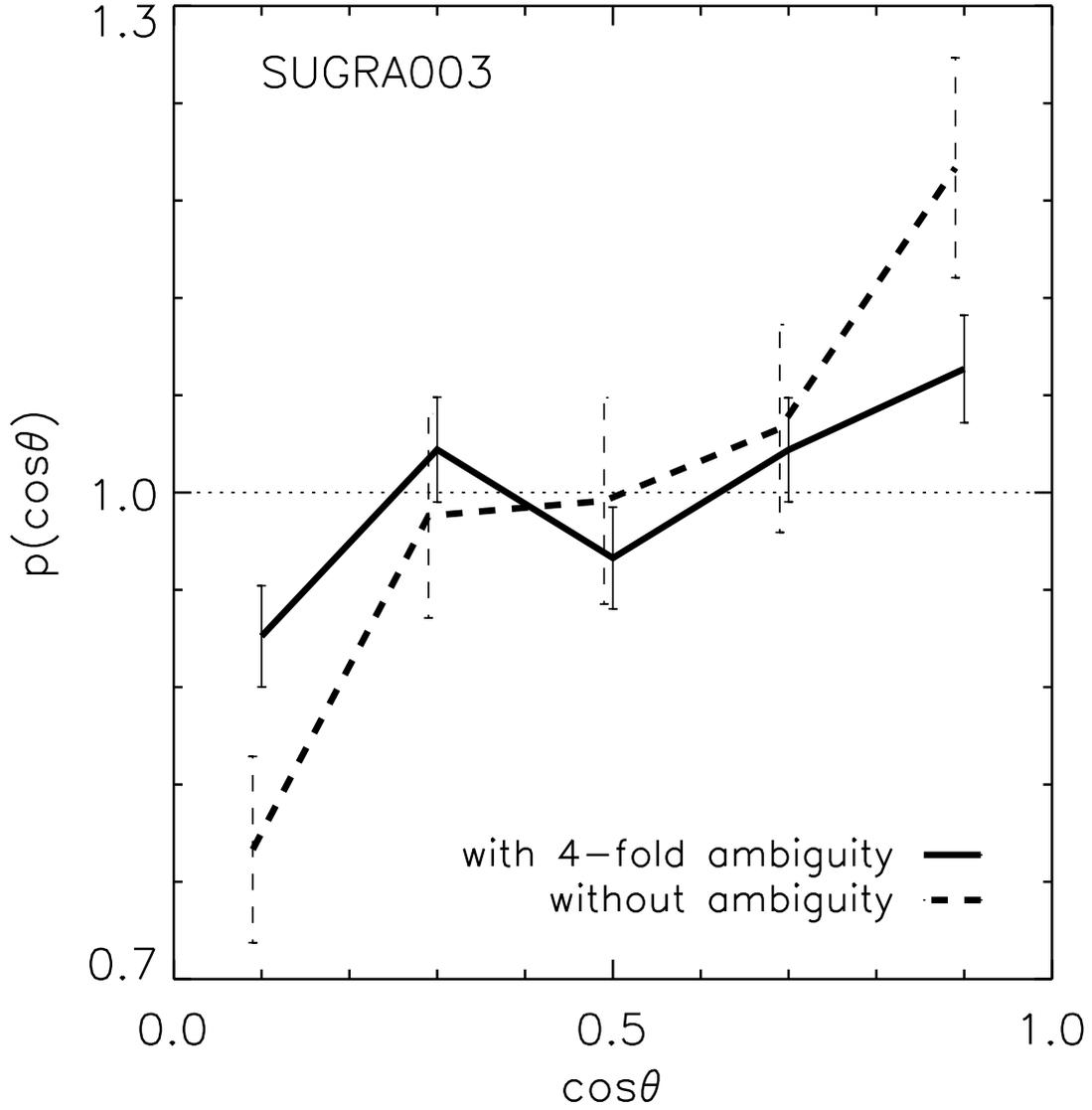}
\caption{Probability density distribution of $\cos\theta$ measured up to 
the four-fold ambiguity for the SUGRA003 model (solid line). The original 
probability density distribution of $\cos\theta$ measured without any ambiguity 
for the same model is also plotted for comparison (dashed line). We offset slightly the 
positions of $\cos\theta$ between the two cases to show the errors more clearly.}
\label{fig:4fold}
\end{center}
\end{figure}

Figure \ref{fig:4fold} plots the newly determined probability density distribution of the 
cosines of the spin alignment angles (solid line) for the SUGRA003 case with the bootstrap 
errors and compare it with the original distribution (dashed line). As can be seen, the 
spin alignment signal becomes weaker, as expected, while the size of the bootstrap 
errors shrinks. Note that although the alignment signal is weaker than the original 
one, it is still statistically significant. We find $\chi^{2}=15.23$ even when the 
four-fold ambiguity is assumed, which still rejects the null hypothesis of $p(\cos\theta)=1$ 
at the $99.8\%$ confidence level. 
In other words, for the SUGRA003 case, the signal of the spin alignment in isolated 
galaxy pairs is so strong and robutst that its statistical significance survives the 
application of the four-fold ambiguity.

\section{DISCUSSION AND CONCLUSION}

In the linear tidal torque theory, the angular momentum of a galactic halo 
originates from the tidal interaction with the surrounding matter at its 
proto-galactic stage \citep{pee69,dor70,white84}. The gas clouds in proto-
galaxies are believed to share the same specific angular momentum with the CDM 
particles provided that they were well mixed with each other at the initial stage 
\citep{FE80}. 
If the galaxies kept the initial memory of the tidal influence, then the spin 
axes in galaxy pairs would align well with each other since both of the two spin 
axes are correlated with the principal axis of the same local tidal field 
\citep{dor70,white84,CT96,LP00,PLS00,porciani-etal02}. In reality, however, the 
galaxies gradually lose their initial tendency of the spin alignments during the 
nonlinear evolutionary processes after decoupling from the Hubble flows 
\citep[e.g.,][]{LE07,lee11}. Henceforth, the degree of the spin alignments in 
galaxy pairs would be determined by the competition between the constructive 
tidal influence and the destructive nonlinear effect.

In cDE models the integrated effect of DE-DM interaction enhance the linear 
perturbations in the matter (gas+CDM) density and velocity fields relative to the 
$\Lambda$CDM case \citep{baldi11b,LB11}. Our original idea was that the enhanced 
linear density and velocity perturbations in cDE models might result in 
augmenting the constructive effect of the initial tidal interaction, helping the 
galaxies keep better the initially induced spin alignments. Here, we have confirmed 
this idea through quantitatively investigating the effect of cDE on the spin 
alignments in isolated pairs of galactic halos identified using the publicly 
available data from the H-CoDECS for the five specific cDE models: EXP001, 
EXP002, EXP003, EXP008e3, SUGRA003 as well as for the $\Lambda$CDM model.

For the case of the $\Lambda$CDM model, no signal of the spin alignment in galaxy 
pairs has been  found, which is consistent with the observational result obtained 
using the pairs of the SDSS late-type galaxies 
\citep[see also][]{cervantes-etal10}. The highest degree of the spin 
alignments in isolated pairs of galactic halos has been found in the SUGRA003 
model where the linear velocity perturbations are enhanced before 
$z_{\rm inv}\approx 6.8$ but lowered after $z_{\rm inv}$,  relative to the 
$\Lambda$CDM case. Here the redshift $z_{\rm inv}$ corresponds to the epoch when 
the cDE bounces on the $\Lambda$-barrier of $P_{\phi}/\rho_{\phi}=-1$, 
inverting its direction of motion 
\citep[see][for a detailed explanation]{sugra003,codecs}. 
Given the peculiar dynamics of the SUGRA003 cDE, the strongest spin alignments 
exhibited by this model can be understood as follows. 
The enhanced velocity perturbations before $z_{\rm inv}$ augment the constructive 
tidal effect of aligning the spin orientations in pairs, while the lowered velocity 
perturbations after $z_{inv}$ diminish the destructive nonlinear effect of erasing 
the tidally induced alignments. 

In fact, bouncing cDE models have recently attracted sharp attentions since   
several observational tensions of the $\Lambda$CDM model have been found to 
be alleviated in bouncing cDE models. 
For instance, \citet{sugra003} have shown that the bouncing cDE accelerates  
the formation of massive clusters and increases their abundance at high redshifts 
without affecting their abundance at low redshifts. 
\citet{LB11} also showed that the pairwise speeds of the colliding clusters 
are significantly enhanced due to the effect of bouncing cDE and thus finding 
a bullet-like cluster is no longer an exceptionally rare event in the SUGRA003 
model. 

It is intriguing to see that our result presented in this Paper 
might be used to rule out the SUGRA003 as a viable cDE model.
The strong signal of spin alignments in pairs of galactic halos predicted by the 
SUGRA003 model is in direct conflict with no observational signal of spin 
alignment in the pairs of the spiral galaxies from the SDSS DR7. What our 
result truly implies, however, is contingent upon how well the observed minor axes 
of the disks in the spiral galaxies are aligned with the directions of the angular 
momentum vectors of their host halos, given that what can be readily measured 
in observations is not the specific angular momentum of a galactic halo but
only the minor axes of its luminous disk. 

According to the recent hydrodynamical simulations, the minor axes of the luminous disks 
are almost perfectly aligned with the angular momentum vectors of the inner halos (where 
the disks are embedded) but only weakly aligned with the angular momentum vectors of 
the entire halos \citep[e.g.,][]{bailin-etal05,hahn-etal10}. 
In spite of these numerical counter-evidences, this alignment issue is still 
inconclusive and related to the long-standing angular momentum problem that the 
disks of the simulated galaxies in a $\Lambda$CDM universe are much smaller in 
size than the observed disks of the spiral galaxies . In other words, the current 
hydrodynamical simulations have failed in producing the extended fast rotating 
disks in spiral galaxies \citep[see, e.g.,][]{BD04,donghia-etal06}. Without 
having a solution to this cosmological angular momentum problem, it is still 
premature to claim that the spin axes of the luminous disks are not so strongly 
aligned with those of the entire halos. 

As a final conclusion, since the degree of the spin alignments in isolated pairs 
of galactic halos is found to depend sensitively on the strength of coupling and 
shapes of the self-interaction potential, it can be in principle used as a unique 
and powerful test of cDE cosmologies, provided that it is understood how well the 
angular momentum vectors of the galactic halos are aligned with the observable 
minor axes of the luminous disks. 

\acknowledgments

I am grateful to a referee for helpful suggestions which helped me improve 
the original manuscript.
I acknowledge the financial support from the National Research Foundation 
of Korea (NRF) grant funded by the Korea government (MEST, No.2011-0007819) 
and from the National Research Foundation of Korea to the Center for Galaxy 
Evolution Research.

\clearpage
\clearpage
\clearpage
\clearpage
\clearpage
\clearpage
\clearpage
\clearpage
\clearpage
\clearpage
\clearpage
\begin{deluxetable}{ccccc}
\tablewidth{0pt}
\setlength{\tabcolsep}{5mm}
\tablecaption{model, \# of the isolated galaxy pairs and masses of the two 
component galaxies in pairs}
\tablehead{model & $N_{pair}$ & $M_{T,\rm med}$ & $m_{\rm 1,med}$ & 
$m_{\rm 2,med}$\\ 
& & $[10^{11}\,h^{-1}M_{\odot}]$ & $[10^{11}\,h^{-1}M_{\odot}]$ & 
$[10^{11}\,h^{-1}M_{\odot}]$} 
\startdata
$\Lambda$CDM  & $948$  & $2.28$ & $1.72$ & $0.35$\\
EXP001  & $914$ & $2.38$ & $1.82$ & $0.34$\\ 
EXP002 & $860$ & $2.35$ & $1.79$ & $0.35$\\  
EXP003  & $774$  & $2.52$ & $1.86$  & $0.34$\\
EXP008e3 & $783$ & $2.53$ & $1.86$  & $0.34$\\
SUGRA003 & $1003$ & $2.33$ & $1.82$ & $0.35$\\  
\enddata
\label{tab:pair}
\end{deluxetable}


\clearpage
\begin{thebibliography}{}
\bibitem[Abazajian et al.(2009)]{sdssdr7} 
Abazajian, K.~N., Adelman-McCarthy, J.~K., 
Ag{\"u}eros, M.~A., et al.\ 2009, \apjs, 182, 543 
\bibitem[Amendola(2000)]{amendola00} 
Amendola, L.\ 2000, \prd, 62, 043511 
\bibitem[Amendola(2004)]{amendola04} 
Amendola, L.\ 2004, \prd, 69, 103524 
\bibitem[Bailin et al.(2005)]{bailin-etal05} 
Bailin, J., Kawata, D., Gibson, B.~K., et al.\ 2005, \apjl, 627, L17 
\bibitem[Baldi \& Viel(2010)]{BV10} 
Baldi, M., \& Viel, M.\ 2010, \mnras, 409, L89 
\bibitem[Baldi et al.(2010)]{baldi-etal10} 
Baldi, M., Pettorino, V., Robbers, G., \& Springel, V.\ 2010, \mnras, 403, 1684 
\bibitem[Baldi \& Pettorino(2011)]{BP11} 
Baldi, M., \& Pettorino, V.\ 2011, \mnras, 412, L1 
\bibitem[Baldi et al.(2011)]{BLM11} 
Baldi, M., Lee, J., \& Macci{\`o}, A.~V.\ 2011, \apj, 732, 112 
\bibitem[Baldi(2011a)]{exp008e3} 
Baldi, M.\ 2011a, \mnras, 411, 1077 
\bibitem[Baldi(2011b)]{baldi11b} 
Baldi, M.\ 2011b, \mnras, 414, 116 
\bibitem[Baldi(2011c)]{codecs} 
Baldi, M.\ 2011c, arXiv:1109.5695 
\bibitem[Baldi(2012)]{sugra003} 
Baldi, M.\ 2012, \mnras, 420, 430 
\bibitem[Baldi \& Salucci(2012)]{BS11} 
Baldi, M., \& Salucci, P.\ 2012, Journal of Cosmology and Astroparticle 
Physics, 2, 14  
\bibitem[Barnes \& Efstathiou(1987)]{BE87} 
Barnes, J., \& Efstathiou, G.\ 1987, \apj, 319, 575 
\bibitem[Bean et al.(2008)]{bean-etal08} Bean, R., Flanagan, 
{\'E}.~{\'E}., Laszlo, I., \& Trodden, M.\ 2008, \prd, 78, 123514 
Benson, A.~J.\ 2005, \mnras, 358, 551 
\bibitem[Brax \& Martin(1999)]{BM99} 
Brax, P.~H., \& Martin, J.\ 1999, Physics Letters B, 468, 40 
\bibitem[Bullock et al.(2001)]{bullock-etal01} 
Bullock, J.~S., Dekel, A., Kolatt, T.~S., et al.\ 2001, \apj, 555, 240 
\bibitem[Burkert \& D'Onghia(2004)]{BD04} 
Burkert, A.~M., \& D'Onghia, E.\ 2004, Penetrating Bars Through Masks of Cosmic 
Dust, 319, 341
\bibitem[Catelan \& Theuns(1996)]{CT96} 
Catelan, P., \& Theuns, T.\ 1996, \mnras, 282, 436 
\bibitem[Cervantes-Sodi et al.(2010)]{cervantes-etal10} 
Cervantes-Sodi, B., Hernandez, X., \& Park, C.\ 2010, \mnras, 402, 1807 
\bibitem[Dalcanton et al.(1997)]{dalcanton-etal97} 
Dalcanton, J.~J., Spergel, D.~N., \& Summers, F.~J.\ 1997, \apj, 482, 659 
\bibitem[Davis et al.(1985)]{FoF02}
Davis, M., Efstathiou, G., Frenk, C.~S., \& White, S.~D.~M.\ 1985,
\apj, 292, 371
\bibitem[de Blok(2010)]{deblok10} 
de Blok, W.~J.~G.\ 2010, Advances in Astronomy, 2010,  
\bibitem[Doroshkevich(1970)]{dor70} 
Doroshkevich, A.~G.\ 1970, Astrofizika, 6, 581 
\bibitem[D'Onghia et al.(2006)]{donghia-etal06} 
D'Onghia, E., Burkert, A., Murante, G., \& Khochfar, S.\ 2006, \mnras, 372, 1525 
\bibitem[Fall \& Efstathiou(1980)]{FE80} 
Fall, S.~M., \& Efstathiou, G.\ 1980, \mnras, 193, 189 
\bibitem[Faltenbacher et al.(2008)]{fal-etal08} Faltenbacher, A., 
Jing, Y.~P., Li, C., et al.\ 2008, \apj, 675, 146 
\bibitem[Hahn et al.(2010)]{hahn-etal10} 
Hahn, O., Teyssier, R., \& Carollo, C.~M.\ 2010, \mnras, 405, 274 
\bibitem[Haynes \& Giovanelli(1984)]{HG84} 
Haynes, M.~P., \& Giovanelli, R.\ 1984, \aj, 89, 758 
\bibitem[Huertas-Company et al.(2011)]{huertas-etal11} 
Huertas-Company, M., Aguerri, J.~A.~L., Bernardi, M., Mei, S., \& 
S{\'a}nchez Almeida, J.\ 2011, \aap, 525, A157 
\bibitem[Jimenez et al.(1997)]{jimenez-etal97} 
Jimenez, R., Heavens, A.~F., Hawkins, M.~R.~S., \& Padoan, P.\ 1997, 
\mnras, 292, L5 
\bibitem[Jimenez et al.(1998)]{jimenez-etal98} 
Jimenez, R., Padoan, P., Matteucci, F., \& Heavens, A.~F.\ 1998, 
\mnras, 299, 123 
\bibitem[Kuzio de Naray \& Spekkens(2011)]{kuzio-etal11} 
Kuzio de Naray, R., \& Spekkens, K.\ 2011, \apjl, 741, L29 
\bibitem[Komatsu et al.(2011)]{wmap7}
Komatsu, E., et al.\ 2011, \apjs, 192, 18
\bibitem[Kroupa et al.(2010)]{kroupa-etal10} 
Kroupa, P., Famaey, B., de Boer, K.~S., et al.\ 2010, \aap, 523, A32 
\bibitem[Lee \& Pen(2000)]{LP00}
Lee, J. \& Pen, U. L. 2000, \apj, 532, L5
\bibitem[Lee \& Pen(2001)]{LP01}
Lee, J. \& Pen, U. L. 2001, \apj, 555, 106
\bibitem[Lee \& Erdogdu(2007)]{LE07}
Lee, J. \& Erdogdu, P.  2007, 671, 1248 
\bibitem[Lee \& Komatsu(2010)]{LK10} 
Lee, J., \& Komatsu, E.\ 2010, \apj, 718, 60
\bibitem[Lee \& Baldi(2011)]{LB11} 
Lee, J., \& Baldi, M.\ 2011, arXiv:1110.0015 
\bibitem[Lee(2011)]{lee11} Lee, J.\ 2011, \apj, 732, 99 
\bibitem[Lucchin \& Matarrese(1985)]{LM85} 
Lucchin, F., \& Matarrese, S.\ 1985, \prd, 32, 1316 
\bibitem[Macci{\`o} et al.(2004)]{maccio-etal04} 
Macci{\`o}, A.~V., Quercellini, C., Mainini, R., Amendola, L., 
\& Bonometto, S.~A.\ 2004, \prd, 69, 123516 
\bibitem[Mainini \& Bonometto(2006)]{MB06} 
Mainini, R., \& Bonometto, S.\ 2006, \prd, 74, 043504 
\bibitem[Mangano et al.(2003)]{mangano-etal03} 
Mangano, G., Miele, G., \& Pettorino, V.\ 2003, Modern Physics Letters A, 18, 831 
\bibitem[Navarro et al.(1996)]{nfw96} 
Navarro, J.~F., Frenk, C.~S., \& White, S.~D.~M.\ 1996, \apj, 462, 563
\bibitem[Paz et al.(2011)]{paz-etal11} Paz, D.~J., Sgr{\'o}, 
M.~A., Merch{\'a}n, M., \& Padilla, N.\ 2011, \mnras, 414, 2029 
\bibitem[Peebles(1969)]{pee69} 
Peebles, P.~J.~E.\ 1969, \apj, 155, 393 
\bibitem[Pen et al.(2000)]{PLS00} Pen, U.-L., Lee, J., 
\& Seljak, U.\ 2000, \apjl, 543, L107 
\bibitem[Peebles \& Nusser(2010)]{PN10} 
Peebles, P.~J.~E., \& Nusser, A.\ 2010, \nat, 465, 565 
\bibitem[Perivolaropoulos(2008)]{puzzle} 
Perivolaropoulos, L.\ 2008, arXiv:0811.4684 
\bibitem[Pettorino \& Baccigalupi(2008)]{PB08} 
Pettorino, V., \& Baccigalupi, C.\ 2008, \prd, 77, 103003 
\bibitem[Porciani et al.(2002)]{porciani-etal02} 
Porciani, C., Dekel, A., \& Hoffman, Y.\ 2002, \mnras, 332, 325 
\bibitem[Ratra \& Peebles(1988)]{RP88} 
Ratra, B., \& Peebles, P.~J.~E.\ 1988, \prd, 37, 3406 
\bibitem[Springel \& Hernquist(2002)]{SH02} 
Springel, V., \& Hernquist, L.\ 2002, \mnras, 333, 649 
\bibitem[Springel(2005)]{gadget2}
Springel, V.\ 2005, \mnras, 364, 1105
\bibitem[Wall \& Jenkins(2003)]{WJ03} 
Wall, J.~V., \& Jenkins, C.~R.\ 2003, Practical statistics for astronomers, 
(Cambridge: New-York) 
\bibitem[Wetterich(1988)]{wetterich88} 
Wetterich, C.\ 1988, Nuclear Physics B, 302, 668 
\bibitem[Wetterich(1995)]{wetterich95} 
Wetterich, C.\ 1995, \aap, 301, 321 
\bibitem[Wintergerst \& Pettorino(2010)]{WP10} 
Wintergerst, N., \& Pettorino, V.\ 2010, \prd, 82, 103516 
\bibitem[Xia(2009)]{xia09} 
Xia, J.-Q.\ 2009, \prd, 80, 103514 
\bibitem[White(1984)]{white84} 
White, S.~D.~M.\ 1984, \apj, 286, 38 
\end{thebibliography}
\end{document}